\documentclass[11pt]{article}
\usepackage{blois,epsfig}
\usepackage[latin1]{inputenc}
\usepackage[OT1]{fontenc}

\def\apj{{\em Astrophys.J.}, }
\def\aj{{\em Astron.J.}, }
\def\aa{{\em Astron.Astrophys}, }
\def\arev{{\em Ann.Rev.Astron.Astrophys.}, }

\def\be{\begin{equation}}
\def\ee{\end{equation}}
\def\bea{\begin{eqnarray}}
\def\eea{\end{eqnarray}}

\begin{document}
\vspace*{4cm}
\title{FROM DARK MATTER TO MOND}

\author{ R.H. Sanders}

\address{Kapteyn Astronomical Institute, University of Groningen\\
Groningen, The Netherlands}

\maketitle\abstracts{MOND-- modified Newtonian dynamics-- may be
viewed as an algorithm for calculating the distribution of
force in an astronomical object from the observed distribution
of baryonic matter.  The fact that it works for galaxies is quite
problematic for Cold Dark Matter.  Moreover, MOND explains or
subsumes systematic aspects of galaxy photometry
and kinematics-- aspects that CDM does not address or gets
wrong.  I will present evidence here in support of these
assertions and claim that this is effectively a falsification of
dark matter that is dynamically important on the scale of galaxies.
 }

\section{Introductory remarks}

Modified Newtonian dynamics, or MOND, was proposed by Milgrom~\cite{mil83}
as an alternative to dark matter.  Over the past 25 years a 
considerable lore has grown up around this idea, and now the very word
seems to provoke strong reactions-- pro or con-- depending upon
ones preconceptions or inclinations.  Here I want to provide
a minimalist definition of MOND-- a definition which is as free
as possible from emotive charge of the idea; therefore, I will avoid
terms like modified inertia or modified gravity.

{\it MOND is an algorithm that permits one to calculate the
distribution of force in an object from the observed distribution
of baryonic matter with only one additional fixed parameter having
units of acceleration.}

This algorithm works very well on the scale of galaxies.  The fact
that it works is problematic for Cold Dark Matter (CDM), because this
is not something that dark matter can naturally do.  Moreover, MOND
explains or subsumes systematic aspects of galaxy photometry
and kinematics-- aspects which CDM does not address or gets 
wrong.  Several of these systematics were not evident at the time
that MOND was proposed, so this constitutes a predictive power going
beyond the ability to explain observations {\it a posteriori}.

I will present the evidence in favor of these assertions,
so this will be a discussion primarily of the phenomenology.  I will,
however, draw the conclusion which to me is also minimal and quite obvious:
standard CDM is falsified by the existence of this successful 
algorithm.

\section{The algorithm and its immediate consequences}

MOND, in its original form, is embodied by Milgrom's simple 
formula: if ${\bf g}$ is the true gravitational acceleration
and ${\bf g_N}$ is the traditional Newtonian acceleration,
then these are related by
$${\bf g}\, \mu(|{\bf g}|/a_0) = {\bf g_N} \eqno(1)$$
where $a_0$ is a new fixed parameter having units of acceleration
and $\mu$ is a function whose form is not specified but which must
have the asymptotic behavior $\mu(x) = 1$ when $x>>1$ 
(the Newtonian limit) and $\mu(x)=x$ when $x<<1$ (the MOND limit).
This means
that the discrepancy between the observed baryonic mass
and the Newtonian dynamical mass should appear below a critical
acceleration.  In fact, for spiral galaxies this is the case;
objects with the lowest measured centripetal acceleration (from
the rotation curves) have the largest implied Newtonian dynamical 
mass-to-light ratios~\cite{sm02}.  The critical
acceleration has a value of $a_0\approx 10^{-8}$ cms$^{-2}$
or about an angstrom per second per second.  This
is within a factor of 10 of $cH_0$-- a coincidence originally 
pointed out by Milgrom~\cite{mil83}.
I might just add 
here that the sort of CDM halos which emerge from cosmic N-body
simulations~\cite{nfw}
do not possess such a natural acceleration scale not withstanding
convoluted arguments to the contrary~\cite{kt}.

This formula has two immediate and significant consequences.
In the limit of low accelerations, the ``true'' gravitational 
acceleration would be given by
$$g=\sqrt{a_0g_N}. \eqno(2)$$
At large distance from a point mass $M$ this would mean that the
effective gravitational attraction is
$$g=\sqrt{GMa_0/r^2}.\eqno(3)$$
If we equate this to the centripetal acceleration we have an asymptotic
rotation velocity given by
$${V_\infty}^4= GMa_0.\eqno(4)$$
In other words, all rotation curves are asymptotically flat and there
exists a mass-rotation velocity relation of the form $M\propto{V_\infty}^4$.

It is often said that MOND was ``designed'' to fit galaxy data, but
Milgrom's proposal
actually predated most of the precise data on galaxy rotation curves--
data such as that shown in Fig.\ 1.
This rotation curve is derived from 21 cm line 
observations of the nearby spiral galaxy NGC 2403
and goes well beyond the bright optical disk~\cite{beg87}.
The dashed and dotted curves are the
Newtonian rotation curves calculated from the observed distribution
of starlight and neutral hydrogen surface density respectively,
and the solid curve is that determined from the Newtonian
force using eq.\ 1.
Here we see that the rotation curve is indeed asymptotically flat, but there
is more to it.  The required mass-to-light ratio in the disk
(M/L=0.9) is entirely reasonable for the population of stars in
this galaxy, and the calculated rotation curve even appears to
reproduce structure in the observed rotation curve in
the inner regions (more on this later).

\begin{figure}[h]
\begin{center}
\psfig{figure=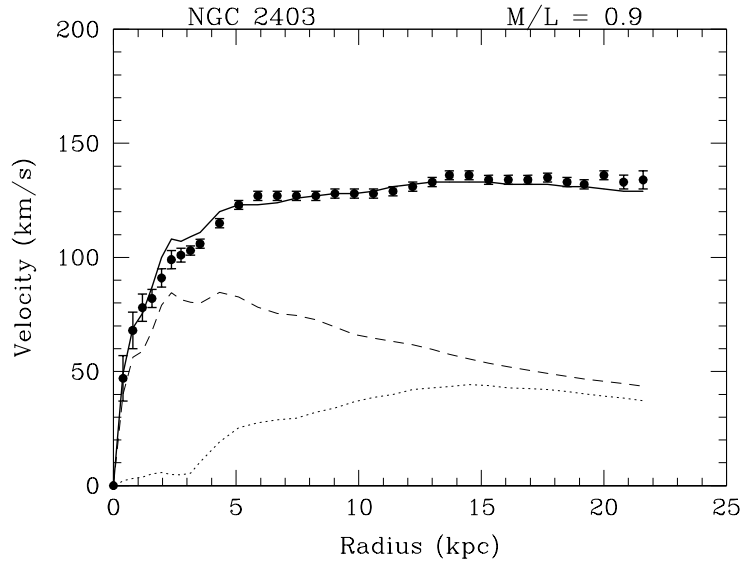,height=80mm}
\caption{The points show the rotation curve of NGC 2403 as 
deduced from 21 cm line observations [6].  The dashed curve is the
Newtonian rotation curve of the stellar component as deduced
from the observed surface brightness distribution with M/L=0.9,
and the dotted curve is the Newtonian rotation curve deduced
from the observed HI surface density distribution.  The solid
curve is that calculated from Milgrom's formula. Here $a_0=10^{-8}$
cms$^{-2}$.}
\end{center}
\end{figure}

The second consequence, the mass-rotation velocity relationship,
forms the basis of the observed Tully-Fisher law-- a correlation
between rotation velocity and luminosity in spiral galaxies.
This observed relation has been converted by McGaugh~\cite{mcg07} to a
baryonic Tully-Fisher relation shown in Fig.\ 2.  Here the stellar 
luminosity is re-expressed as a stellar mass, and the directly
observed mass of
gas, important in dwarf galaxies, is also included.
The solid line is not a fit but the expectation from MOND with
$a_0 = 10^{-8}$ cms$^{-2}$.
Thus MOND subsumes the baryonic Tully-Fisher law: it exhibits
the observed slope, and the correct zero-point, and the relationship
is exact (the only scatter is observational).

\begin{figure}[h]
\begin{center}
\psfig{figure=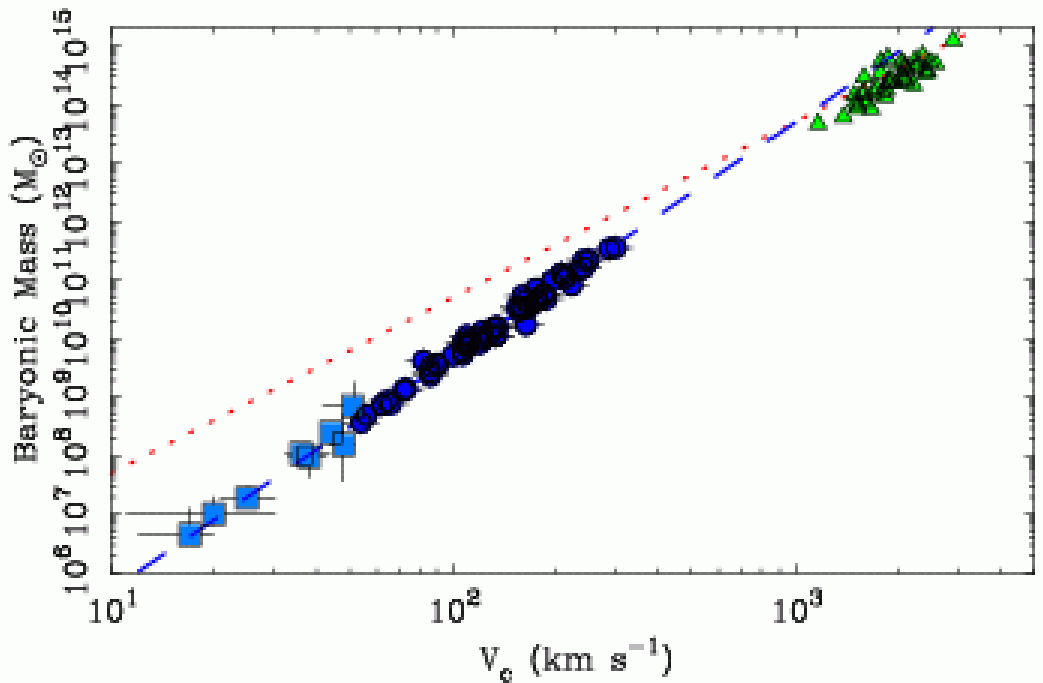,height=68mm}
\caption{The baryonic Tully-Fisher relation derived by McGaugh [7].  
This is
the the baryonic mass plotted against the rotation
velocity for a sample of spiral galaxies. The squares are
dwarf galaxies where gas makes a dominant contribution
to the baryonic mass. The triangular points are the deduced circular
velocity for clusters of galaxies.  The solid line is the MOND relation
and the dotted line is that predicted from CDM simulations.}
\end{center}
\end{figure}

In the context of dark matter, the baryonic Tully-Fisher relation
would be a correlation between the tiny bit of remaining
baryonic mass (after blowout)
in the central regions of a vast halo and the circular velocity
established by that halo.
How does CDM explain this near perfect correlation?  In CDM cosmological
simulations all halos at a given cosmic time have about the same density
(with considerable scatter).  If we combine this with the Newtonian virial
theorem we find that $M\propto V^3$ (with considerable 
scatter)~\cite{sn99}.  This,
in itself, is inconsistent with observations as is illustrated in Fig. 2
by the dotted line.  To bring CDM expectation in line with the
observations, protogalaxies must loose a fraction of their baryonic
mass which is greater for lower mass galaxies.  And it must do this
in a manner which {\it reduces} the scatter.  There has been a great deal
of work trying to accomplish this (with so-called ``semi-analytic
galaxy formation'' programs), but it remains an exercise characterized
by parameter tuning in order to achieve the desired result.  It is
characteristic of the CDM paradigm with respect to explaining
global scaling relations:  The scaling relations arise from
aspects of galaxy formation.  The pure N-body simulations do not
explain the observed relations, so advocates fall back on ``complicated
baryonic physics'' or ``gastrophysics'' to push the expectations to
conform to observations.  With MOND, scaling laws like Tully-Fisher
do not result from the details of galaxy formation but from existent dynamics. 
The same is true of the general trends which I enumerate below.

\section{General trends embodied by MOND}

\noindent 1.  The MOND acceleration may be written as a surface density
$$\Sigma_c=a_0/G\eqno(5)$$
Numerically this corresponds to 860 ${M_\odot}/pc^2$, or, if
we assume a mass-to-light ratio of one to two in solar units in the
blue band, to
a surface brightness of $\mu_B\approx 22$ mag/arcsec$^2$.
When an astronomical system has a surface brightness comparable or
larger than this value, it implies that $\Sigma\geq\Sigma_c$ and
the object is in the high-acceleration or Newtonian regime.  Then
there should be no significant discrepancy between the
Newtonian dynamical mass and the detectable baryonic mass,
at least not within the bright regions of the system.  
Examples of high surface brightness systems are globular star clusters
and luminous elliptical galaxies.  It is well known that the 
Newtonian dynamical mass-to-light ratio in globular clusters 
is entirely consistent with normal stellar populations-- there
is no evidence for dark matter~\cite{peal89}.  
Furthermore, MOND predicts a ``dearth
of dark matter'' in luminous elliptical galaxies,
 a dearth that has been
recently confirmed~\cite{romeal03,ms03}. 
To falsify MOND, one need only find high
surface brightness systems which require dark matter within
the optically visible image.

On the other
hand, for low surface brightness systems $\Sigma<\Sigma_c$ and the object is
in the low acceleration or MOND regime.  This means that in
such systems there should be a large discrepancy
between the Newtonian mass and the baryonic mass.  This is certainly
true.  In dwarf spheroidal galaxies, the very faint low surface brightness
companions of the Milky Way, there is indeed a very large discrepancy
~\cite{mat}--
in traditional language-- the systems are dominated by ``dark matter.''
Since Milgrom's original papers, many low surface brightness
disk galaxies, with well measured rotation curves, have been discovered.
These objects, without exception, exhibit a large discrepancy within
the visible disk~\cite{mcgdb}.  

\medskip

\noindent 2. It has been known for many years that rotationally
supported Newtonian disks tend to be unstable~\cite{op}; in fact, 
this was an
original motivation for pressure supported dark halos around 
spiral galaxies. But, in the context of MOND, 
disks with surface densities below $\Sigma_c$
are not Newtonian and thus stable.  Therefore, MOND implies that there
should be an upper limit to the surface density or surface brightness
of rotationally supported disks.  There is:  it is called the
Freeman limit and is roughly 22 mag/arcsec$^2$, corresponding
to $\Sigma_c$.

\begin{figure}[h]
\begin{center}
\psfig{figure=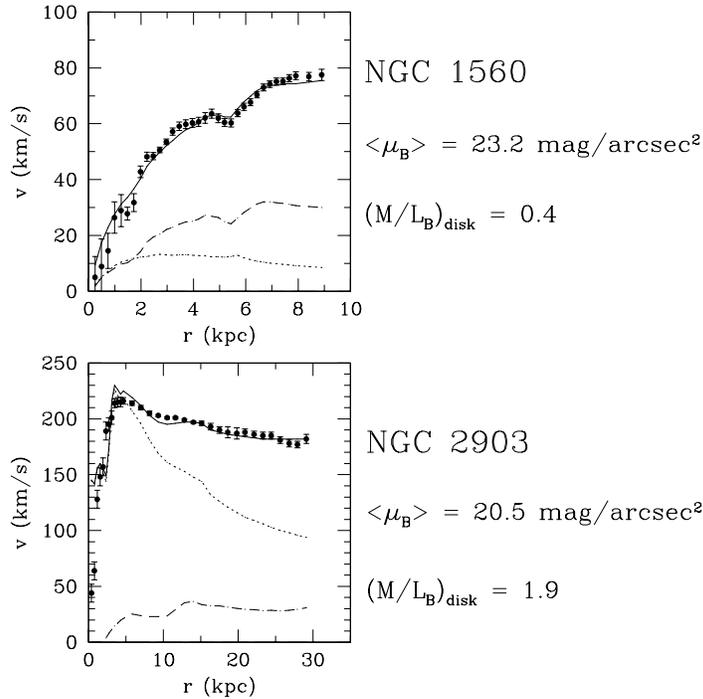,height=100mm}
\caption{Observed rotation curves of a low surface brightness 
(Broeils [15])and
a high surface brightness galaxy (Begeman [6]). Here the dotted curve 
is the
Newtonian rotation curve of the stellar component and the dashed
curve for the gas.  The solid curve is the MOND rotation curve.
The mean surface brightness and the implied mass-to-light ratios
are indicated.}
\end{center}
\end{figure}

\medskip
\noindent 3.  Low surface brightness disk galaxies are in the deep
MOND regime.  Therefore the rotation curve should slowly rise to
the asymptotic value which is the maximum rotational velocity.
On the other hand, high surface brightness disks are in the
Newtonian regime.  Therefore the rotation curves, after an abrupt
rise, should slowly decline in a near Keplerian fashion to the
final asymptotic velocity.  These predictions have been subsequently
confirmed by observations of LSB and HSB disk galaxies; examples
are shown in Fig.\ 3.  

\medskip

\noindent 4.  Not all galaxies are rotationally supported disks;
there are also galaxies held up against gravity by the random
motion of the stars-- pressure supported systems.  In fact, 
not only galaxies fall into this category but there are pressure
supported systems ranging from globular star clusters 
(with $10^5$ M$_\odot$) to the great clusters of galaxies with
a baryonic content consisting primarily of hot gas having a total
mass approaching $10^{14}$ M$_\odot$. A fair approximation
for such systems is the isothermal sphere-- an object with
a constant velocity dispersion.  When we solve the equation
of hydrostatic equilibrium for such systems, using Milgrom's
formula, we find that unlike their strictly Newtonian counterparts,
isothermal spheres have finite mass.  In the inner high acceleration regime,
the density falls as $1/r^2$ as in a Newtonian isothermal sphere,
but then at distance where the acceleration approaches $a_0$
($r=\sqrt{GM/a_0}$) the sphere rather abruptly truncates (density
$\rightarrow 1/r^4$).
This means that the average internal acceleration in
MOND isothermal spheres should be near $a_0$.  

\begin{figure}[h]
\begin{center}
\psfig{figure=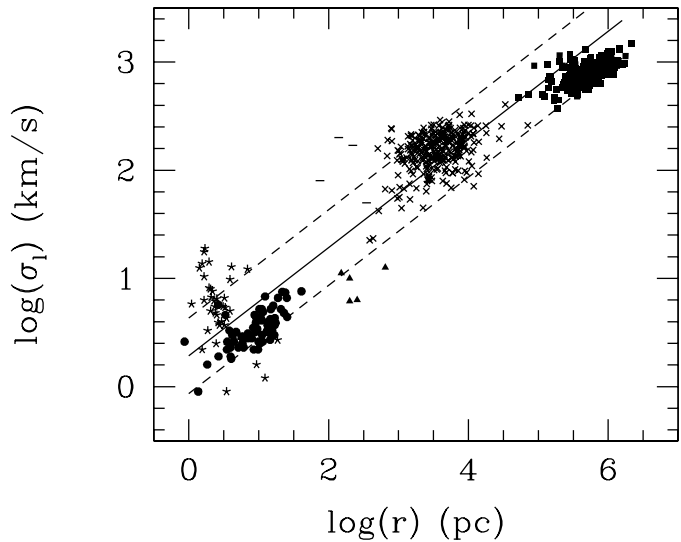,height=80mm}
\caption{A log-log plot of the velocity dispersion in hot
stellar systems against the characteristic size for different
classes of objects.
The star points are globular clusters, the solid round points are
massive molecular clouds in the Galaxy, the crosses are luminous
elliptical galaxies, the triangles are dwarf spheroidal galaxies, the
squares are X-ray emitting clusters of galaxies.  The solid line 
corresponds to $\sigma^2/r = a_0$ and the dashed lines show a factor
of 2.2 variation on each side.  
The references to the relevant observational papers
may be found in [2].}
\end{center}
\end{figure}

Fig.\ 4 shows
the observations~\cite{sm02}.  This is log-log plot of velocity
dispersion vs. size for systems spanning many orders of magnitude
from sub-galactic to super-galactic systems
(the identity of the systems is noted in the caption).  The straight
line is not a fit but rather the locus of $\sigma^2/r=a_0$.
We see that the internal acceleration of these systems all lie
with in a factor of a few of $a_0$.  How does dark matter accomplish 
this?

There is another prediction with respect to 
pressure supported systems originally pointed out by Milgrom.
It is an easy matter to show, directly from the hydrostatic
gas equation that MOND implies that the mass and the velocity
dispersion are related as
$${M\over{10^{11}M_\odot}}\approx {\Bigl[{\sigma\over{100\,km/s}}}\Bigr]^4
\eqno(6)$$
This relation is not so precise as the Tully-Fisher law because the
scaling is quite sensitive to deviations from an isothermal state or
isotropy of the velocity distribution.  It forms the basis of the
Faber-Jackson relation for elliptical galaxies-- an observed
$L\propto \sigma^4$ correlation.  But it goes beyond this; it 
applies to any pressure supported, near isothermal system,
and it tells us that an object with a velocity dispersion of
5 km/s will have a mass of about $10^5$ $M_\odot$ (a globular
cluster), or that an object with a velocity dispersion of
100 km/s will have a galaxy mass, or an object with a velocity
dispersion of 1000 km/s will have a mass ($\approx 10^{14}$ $M_\odot$)
of a cluster of galaxies.

\section{Rotation curves}

Certainly the most remarkable aspect of MOND is its ability
to predict the form of rotation curves from the observed
distribution of baryonic matter.  The procedure has
been discussed many times before~\cite{sm02}.  We assume that the
stellar mass is traced exactly by the surface brightness
distribution (in the near-infrared preferably).  We then
include the observed distribution of gas (which can
make a significant contribution in low mass galaxies), assume
that the stars and gas are distributed in a thin disk (apart
from some spiral galaxies which  possess a significant luminous
bulge or central spheroidal component), apply the traditional
Poisson equation to determine the Newtonian force ($g_N$) and
finally use the MOND formula to calculate the ``true'' gravitational
force.  There is generally one free parameter in all of this--
the mass-to-light ratio of the visible disk which is adjusted to
achieve the best fit.

The results have been impressive, particularly considering all the
intrinsic uncertainties in galaxies (warps, bars, distance uncertainties,
effects of companions).  In about 90\% of the 100 or so rotation curves
considered, MOND produces a very reasonable version of the observed
rotation curve. Moreover, the implied mass-to-light ratios of the
stellar components are not only
sensible, but in complete agreement with population
synthesis models.

\begin{figure}[h]
\begin{center}
\psfig{figure=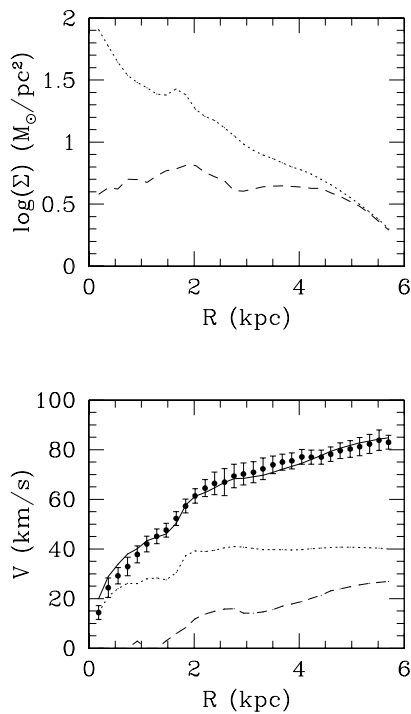,height=100mm}
\caption{Top:  the mass surface density distribution in stars and
gas (dotted and dashed curves) as a function of radius for the
low surface brightness dwarf, UGC 7524.  Bottom:
the corresponding Newtonian 
and the MOND rotation curves (dotted, dashed, solid).  
The points are the observed curve. Observations from ref. [15].}
\end{center}
\end{figure}

I have already shown three examples of observed rotation curves
compared to the curve calculated with the MOND formula using
the observed distribution of detectable baryons (Figs.1 \& 3).
I show one more in Fig.\ 5 because it illustrates very well a point that
I wish to emphasize.  UGC 7524 is a dwarf low surface brightness
galaxy~\cite{swat}.  In the top figure I show the logarithm of the surface
density in stars and gas as a function of radius (the stellar 
surface density is determined from the surface brightness
distribution with MOND value of M/L=1.6).  In the bottom
figure I show again the observed rotation curve (points),
the Newtonian rotation curves of stars and gas, and MOND
rotation curve.  We see that, for both stars and gas, there is
a enhancement in the surface density between 1.5 and 2.0 kpc, and
of course, there is a corresponding feature in the Newtonian
rotation curves.  But we see that there is also a feature
at this position in the total rotation curve, even though there
is a significant discrepancy between the Newtonian and 
detectable mass.  The total rotation curve perfectly reflects
details in the observed mass distribution even though the object
is ``dominated by dark matter'' in the inner regions.

This is an empirical point emphasized repeatedly by Sancisi~\cite{sanc}:
{\it For every feature in the surface brightness distribution
(or gas surface density distribution) there is a corresponding
feature in the observed rotation curve (and vice versa).}
I would add that with dark matter this seems rather unnatural.  How is
it that the dark matter distribution could match so perfectly
the baryonic matter distribution?  But with MOND, it is
expected.  What you see is all there is!

\section{Concluding remarks}

Although eq.\ 1 predicts the detailed distribution of force in galaxies from
the observed distribution of baryonic matter, it appears to break down
in clusters of galaxies.  Applying the MOND formula in the hydrostatic
gas equation, we find that, for X-ray emitting clusters, MOND reduces
the mass discrepancy by a factor of two, but there still remains a factor
of two or three more mass than is directly observed in hot gas and
stars in galaxies.  Formally, this is not a falsification
because we may always find more mass in clusters (it would be a 
falsification if
MOND predicted {\it less} mass than is observed), but this is seen
by some as devastating for a proposed alternative to dark matter.

I take quite the opposite point of view.  The existence of an
algorithm which precisely predicts the force in galaxies from the
observed distribution of baryonic matter is devastating for 
dark matter which clusters on the scale of galaxies, CDM.
In fact, it constitutes a falsification.  To explain the
MOND phenomenology with dark matter would require an
intimate dark matter-baryon coupling which is totally at odds
with the proposed nature of CDM.  Baryons behave quite differently
from CDM: they dissipate and collapse to the center of a system;
they are blown out by supernovae; they are left behind in
collisions between galaxies or clusters of galaxies.
The intimate connection of dark matter and baryons implied
by the phenomenology of rotation curves is incomprehensible
in terms of CDM.

In the context of CDM, global scaling relations, such as the 
Tully-Fisher or Faber-Jackson relation, have their
origin in aspects of galaxy formation.  Yet, galaxy formation,
as emphasized by Milgrom~\cite{mil08}, is quite a haphazard process with each
galaxy having its own unique history of formation, interaction,
and evolution.  It is difficult to imagine that the
ratio of baryonic to dark mass would be a constant in galaxies,
or even vary systematically with galaxy mass.  And yet this is
required, in a very precise way, to explain the baryonic Tully-Fisher
relation-- an exact correlation between the baryonic mass and the
asymptotic rotation velocity which supposedly is a property
of the dark matter halo.
Any initial intrinsic
velocity-mass relation of proto-galaxies would surely be erased in
the stochastic process of galaxy formation.  To believe that
vague processes such as ``feedback'' or ``self-regulation'' can
restore even tighter correlations is equivalent to faith in the 
tooth fairy.

And finally, there is the ubiquitous appearance of $a_0\approx cH_0$.
How do CDM halos, which embody no intrinsic acceleration scale, account
for the facts that $a_0$ is the acceleration at which the discrepancy
appears in galaxies, that $a_0$ determines the normalization of the
Tully-Fisher relation for spiral galaxies and the Faber-Jackson
relation for hot systems, that $a_0$ is the characteristic internal
acceleration
of spheroidal systems ranging from sub-galactic objects to
clusters of galaxies, that $a_0$ defines a critical surface brightness
below which the discrepancy is present.

This body of evidence cannot be ignored and constitutes a profound case
against CDM.  Moreover, this phenomenology implies that there is
something essentially correct about MOND.  Although I have avoided
the subject here, the implications are far-reaching.
\section*{Acknowledgments}  I am grateful to Moti Milgrom and
Stacy McGaugh for comments on this presentation.

\end{document}